\documentclass[a4paper,11pt,superscriptaddress,altaffilletter,lengthcheck,tightenlines,showpacs,showkeys]{article}
\usepackage{latexsym}
\usepackage{feynmf}		
\usepackage{graphicx}
\usepackage{epstopdf}
\usepackage[dvipdf]{epsfig}
\usepackage{color}

\def\ba{\begin{eqnarray}}
\def\ea{\end{eqnarray}}

\def\lb{\label}
\def\be{\begin{equation}}
\def\ee{\end{equation}}

\begin{document}
\title{Perturbations in some models of tachyonic inflation}


\author{Iv\'an E. S\'anchez G.  \thanks{Departamento de F\'isica, FCEyN and IFIBA, Universidad de Buenos Aires, Buenos Aires, Argentina 
isg.cos@gmail.com and isg@df.uba.ar.}  and Osvaldo P. Santill\'an \thanks{Departamento de Matem\'atica, FCEyN, Universidad de Buenos Aires, Buenos Aires, Argentina
firenzecita@hotmail.com and osantil@dm.uba.ar.} }
\date {}
\maketitle 

\begin{abstract}
In the present work an inflationary tachyon field model of the early universe is considered. Several cosmological effects produced by a particular potential in this tachyonic era are studied, under the approximation of slow-roll inflation. In particular, the evolution of the spectral index $n_s$ with time is obtained. The equations for the cosmological scalar perturbations are analytically solved in order to show that the power spectrum for small $k$ values is $P_{\zeta}\sim 1/k^{\frac{1}{2}+\nu_2}$, where $\nu_2$ depends on the barotropic index $\gamma_0$. For large $k$ values we find that the power spectrum is well approximated by the standard inflation model. Additionally, the three-point correlation function is calculated in order to get the primordial non-Gaussianity of the perturbation. The result is that $f_{NG} \simeq 0$ so the non-gaussianities generated by this tachyon field are negligible.
 \end{abstract}
\vskip 1cm




\section{Introduction}
The cosmological inflation could explain various problems of the standard Big Bang cosmological model \cite{Staro}. This scenario not only explains the homogeneity and flatness of the universe, but also provides a mechanism to create the primordial inhomogeneities required for structure formation. Quantum fluctuations in the microscopic inflationary region, magnified to cosmic size, become the seeds for the growth of structure in the Universe \cite{Lyth}. The simplest inflationary model is described by a scalar field which slowly rolls down its potential. 

One of the scalar fields which can be responsible for the early time inflation in the history of the universe is the tachyon field \cite{TaS}-\cite{Nozari}. Such a scalar field is most natural in the context of string theory, particularly associates with D-branes \cite{TaS}, where the inflaton is an open string mode describing the brane position in the extra dimensions \cite{Dvali}. This field may be responsible for early time inflation in the history of the universe \cite{FE}, \cite{Sami}, due to tachyon condensation near the top of the effective scalar potential \cite{TaS}, and also can be considered as a dark energy in the late time \cite{Pad},\cite{Abramo}-\cite{SanSan}.

An exotic matter with negative pressure namely, a tachyon coming from string theory, also has been widely applied in cosmology. In \cite{Gibbons}, the tachyon condensate starts to slowly roll down the potential, and an universe dominated by this field evolves smoothly from a phase of accelerated expansion to an era dominated by a non-relativistic fluid. These features show that tachyon fields may provide suitable candidates to realize initial inflation and may generate an important contribution to the density perturbation and non-gaussianity \cite{Lyth1}. This is the reason why we consider a tachyon field for our purposes. An extended version of the tachyon Lagrangian was found in \cite{LuisTE} and \cite{ISG}.

In a scalar field cosmology the potential plays a crucial role in obtaining inflation. The present work considers a hyperbolic function of the field as the potential, similar to the one used in \cite{Malda}, \cite{Crema}, \cite{Leblond}. The potential we have chosen for the tachyon field leads to an effective equation of state which interpolates between a nearly dust era at early times and a de Sitter stage at late times.  

It should be remarked that there exist some studies of inflation in Dirac-Born-Infeld (DBI) models and their non gaussianties \cite{silver}, \cite{silver2}. These models are  string theory inspired. However, the models we are working with are considerably different than the ones studied in that references.

Several stability issues of the tachyon models described above has been studied. However, the analysis done yet is focused for the classical homogeneous and isotropic solution. Therefore it may be of interest to study perturbation issues related to these models. There may be the case that two models seem to be equivalent at the level of the isotropic and homogeneous solutions, but which differs considerably when perturbations are considered. In a previous publication the authors did not considered inflationary aspects of this tachyon fields. They have shown that this models do not describe correctly observed expansion history and the structure formation of the universe \cite{SanSan}. The aim of the present work is to show that, even that drawback, this models are well suited for inflationary purposes. It is shown  in particular that the two point correlation function for the fluctuations for the tachyon model imitates very well a gaussian spectrum, except for wavelengths larger than the actual Hubble radius. In addition the three point function, which vanish for the gaussian case, is also approximately zero in the present case as well. This suggest that this model can imitate very well the standard inflation cosmological model.

The organization goes as follows. Section 2 contains the main aspect of the tachyon model under consideration. In section 3 the main aspects of the slow roll approximation and the power spectrum of the primordial fluctuation in inflationary stages are briefly discussed, and the differential equation that describes the perturbations is reviewed. In addition the two point correlation function for the fluctuations is calculated. The calculation is not exact since the authors were not able to solve the equation for the perturbations explicitly. All the approximations done for this calculation are described explicitly.  The resulting power spectrum imitates very well a gaussian one.  However, a further check is needed in order to understand wether or not the approximations done are correct. For this reason, in section 4  the three point correlation function for the perturbations is calculated, and shown to be negligible for this model, an aspect which imitates a gaussian spectrum. This partially confirms the calculation done in section 3. Section 5 contains a discussion of the results.

\section{Tachyon Field Equations }

The tachyon scalar field $\phi$ to be considered below is described by the following generic lagrangian 
\begin{equation}
L=V(\phi)\sqrt{1-\partial_\mu \phi \partial^\mu \phi}. 
\label{lagran}
\end{equation}
Here $V(\phi)$ is its scalar potential. The generic expressions for the tachyon pressure and  energy density are
\begin{equation}
\label{3}
\rho_\phi={V\over \sqrt{1-\dot\phi^2}},\qquad p_\phi=-{V\sqrt{1-\dot\phi^2}}.
\end{equation}
By taking the tachyonic density Eq. (\ref{3}) into account the modified Friedmann equation can be written as
\begin{equation}
\label{EERS1}
3H^2=\frac{V}{\sqrt{1-\dot{\phi}^2}}+\Lambda.
\end{equation}
The equation of motion for tachyon field $\phi$, derived from the lagrangian (\ref{lagran}) for an isotropic and homogeneous background is
\begin{equation}
\label{kg}
\ddot\phi+3H\dot\phi(1-\dot{\phi}^2)+\frac{1-\dot{\phi}^2}{V}\frac{dV}{d\phi}=0,
\end{equation}
where $H=\dot{a}(t)/a(t)$ is the Hubble parameter and $a(t)$ is the cosmic scale factor. The spatial curvature $k$ has been assumed to vanish. The dot denotes the derivative with respect to coordinate time $t$.
Here $V(\phi)$ is the scalar potential. In the following the potential 
\begin{equation}
\label{Pl}
V(\phi)=\frac{\Lambda\sqrt{1-\gamma_0}}{\sinh^{2}\frac{\sqrt{3\gamma_0\Lambda}}{2}\phi},
\end{equation}
will be considered. This potential has a singularity at $\phi=0$, but it reasonably fits some appearing in bosonic string theory compactifications \cite{Malda} for $\phi\neq 0$.

The  equation of state for this model may be parameterized as a barotropic relation $p_\phi=(\gamma-1)\rho_\phi$ with a time dependent barotropic index $\gamma(t)$.
The Einstein equations together with (\ref{kg}) result in the following equation for the evolution of this index
 \be\lb{6}
 \dot\gamma=2(\gamma-1)\left(3H\gamma+{\dot V\over V}\right).
 \ee
The requirement for stability for the solutions is that the index $\gamma$ tends to a constant value $\gamma_0$ at asymptotic times $t\to \infty$. This condition, combined with (\ref{6}) gives
 \be
 {\dot V\over V}\simeq-3\gamma_0{\dot a\over a}, \qquad\hbox{so that}\qquad V\simeq{V_0\over a^{3\gamma_0}},
 \ee
 and the following asymptotic value for the energy
 \be\lb{7}
 \rho_\phi={V_0\over\sqrt{1-\gamma_0}}\,\, a^{-3\gamma_0}.
 \ee
 The equation (\ref{6}) can be rewritten by use of ({\ref{7})  in the following form
 \be\lb{9}
 \dot\gamma=6H(\gamma-1)(\gamma-\gamma_0).
 \ee
The solution of the differential equation (\ref{9}) for the case where $\gamma_0\neq1$ is
\be
\gamma={\gamma_0+c_1a^{6(\gamma_0-1)}\over 1+c_1a^{6(\gamma_0-1)}},
\ee
where $c_1$ is an integration constant. It can inferred from the above equation that $\gamma$ tends asymptotically
 to $\gamma_0$ once $\gamma_0<1$. On the other hand, when $\gamma_0=1$ the solution of (\ref{9}) reads
 \be
 \gamma=1-{1\over c_2+6\log a},
 \ee
with $c_2$ denoting an integration constant. According to the last equation the barotropic index
tends to $\gamma_0=1$ for large values of $a$.

The task of solving the Eqs. (\ref{EERS1}) and (\ref{kg}) may be complicated depending on the functional form of $V(\phi)$ and the dependence of the field with time, usually requires numerical integration.
For the specific potential choice (\ref{Pl})  this can be achived by considering a linear dependence of the field in time, which is a good approximation in a de Sitter background and in the slow roll approximation \cite{Cook}, 
\begin{equation}
\label{Ft}
\phi=\phi_{0}t, \qquad \dot{\phi}^{2}=\phi_{0}^{2}=\gamma_0,
\end{equation}
for which
\begin{equation}
\label{aL}
a=a_0\left[\sinh\frac{\sqrt{3\gamma_{0}^{2}\Lambda}}{2}t\right]^{2/3\gamma_0},
\end{equation} 
with $a_0=1$ for the scale factor today. 

The Hubble parameter $H$ may be expressed in terms of time and the scale factor. This gives that
\begin{equation}   
\label{HLa}
H^2=\frac{\Lambda}{3}\coth^2\frac{\sqrt{3\gamma_0^2\Lambda}}{2}t, \qquad  H^{2}=\frac{\Lambda}{3}\frac{1+a^{3\gamma_0}}{a^{3\gamma_0}}.
\end{equation}


The main task of this paper is to analyse the power spectrum for fluctuations around this background. The next sections are devoted to this task.

\section{Cosmological scalar perturbations}
 \subsection{Slow roll approximation}
For a tachyonic inflationary model the slow roll parameters, which are defined as $\epsilon=-\dot{H}/H^2$ and $\eta=-\ddot{H}/H\dot{H}$ \cite{Nozari}, take the following form
\begin{equation}
\epsilon=\frac{3}{2}\gamma_0\frac{1}{\cosh^2\frac{\sqrt{3\gamma_{0}^{2}\Lambda}}{2}t}, \qquad \eta=\sqrt{3\gamma_{0}^{2}\Lambda}.
\end{equation} 

The identification of one parameter in the topic of scalar perturbation in cosmology is of particular interest. This is the scalar spectral index $n_s$ , which is related to the power spectrum of the perturbations $\mathcal{P_R}$, through the relation $n_s-1 = d\ln \mathcal{P_R}/d\ln k$. In this model and under slow-roll approximation, it becomes
\begin{equation}
\label{SpecInd}
n_s\simeq 1-6\epsilon+2\eta.
\end{equation}

In Fig. (\ref{F1}) we show the spectral index $n_s$ as a function of time for different values of the barotropic index. For small values of the barotropic index, for example with $\gamma_0=0.004$, the spectral index at a initial value of time is $n_s=0.964$, and the function increase slowly for a inflation period of time, that is in agreement with the data released by Planck Mission \cite{Planck2015}.

\begin{figure}[hbt]
\begin{center}
\includegraphics[width=8cm]{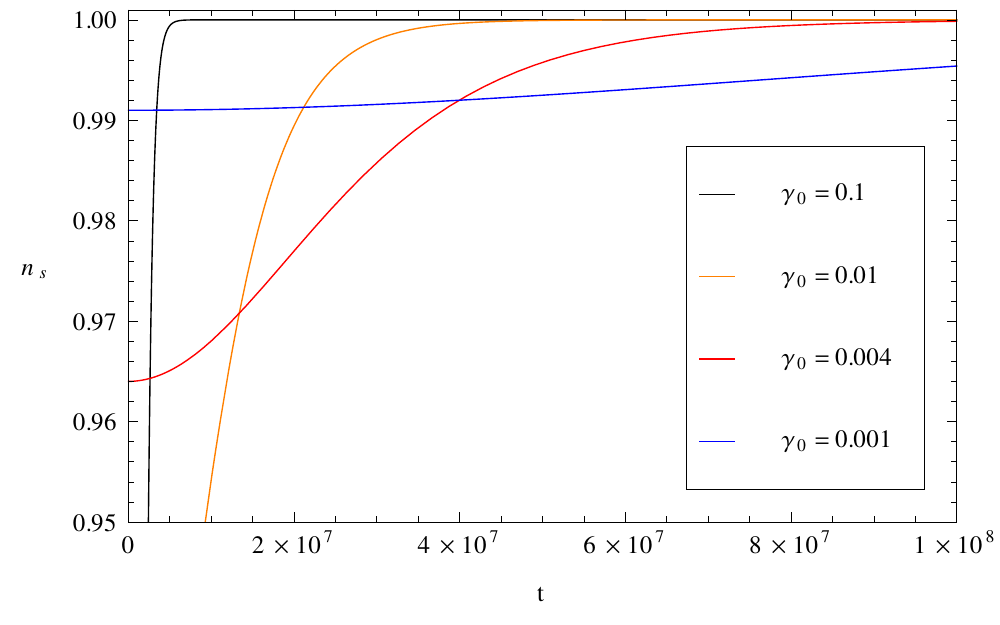}
\caption{\scriptsize{Plot of $n_s(t)$, using the barotropic indexes $\gamma_0=0.1$, $\gamma_0=0.01$, $\gamma_0=0.004$ and $\gamma_0=0.001$.}}
\label{F1}
\end{center}
\end{figure}

The number of e-folds in the slow roll approximation is then given by
\begin{equation}
N(t)=\int_{t_0}^tHdt=\frac{2}{3\gamma_0}\ln \left(\sinh \frac{\sqrt{3\gamma_0^2\Lambda}}{2}t\right)_{t_0}^t,
\end{equation}
where $t_0$ is the initial value of the cosmic time when inflation begins and $t$ is the time value when the inflation ends.

The inflationary phase takes place whenever $\ddot{a} > 0$ which is proportional to $\epsilon < 1$. Consequently, in our model $t>(2/\sqrt{3\gamma_0^2\Lambda}) {\rm arccosh} \sqrt{\frac{3\gamma_0}{2}}$, is the necessary and sufficient condition of inflation. Also, it can be assumed that inflation begins when $\epsilon = 1$, which yields to $t_0=(2/\sqrt{3\gamma_0^2\Lambda}) {\rm arccosh} \sqrt{\frac{3\gamma_0}{2}}$, as the time in which inflation starts. 

 \subsection{Defining equations}
The perturbations around the classical tachyon background solution $\phi(t)$ just described will be described in terms of the methods of reference \cite{Garriga}-\cite{MukLibro}. Following their procedure, the tachyon  $\phi$ 
is decomposed as a sum of a classical solution $\phi(t)$ and a quantum fluctuation $\alpha(x,t)$ as
\be\lb{fluc}
\phi\to \phi(t)+\alpha(x,t).
\ee
The homogeneous and isotropic solution $\phi(t)$ is considered as a classical field, while the fluctuations $\alpha(x,t)$ are quantum fields. The presence of the external classical field $\phi(t)$ creates quantum perturbations $\alpha(x,t)$ in analogous manner as the Schwinger effect of pair creation. Thus, at least theoretically, such scenario can cast inhomogeneity growth from an homogeneous and isotropic universe and may give rise to structure formation. 

The inhomogeneities for the field $\phi(t)$ are sources for perturbations for the metric.  The general form for this perturbations is
\begin{equation}
\label{ds}
ds^2=\emph{a}^2(\tau)\left[(1+2\Phi)d\tau^2-(1-2\Psi)\delta_{ij}dx^{i}dx^{j}\right],
\end{equation}
where $\emph{a}(\tau)$ is the scale factor depending on the conformal time, defined as $d\tau=dt/a$. As can be seen from (\ref{ds}), the quantities $\Phi$ and $\Psi$ can be interpreted as the amplitudes of the metric perturbations in the conformal-Newtonian coordinate system. Since, as discussed in the previous section, the equation of state is of the form $p=(\gamma -1)\rho$, it follows that the non diagonal space-space components of the energy-momentum tensor vanishes. In such situations $\Phi=\Psi$ in  (\ref{ds}) and the distance element becomes
\be\lb{pe}
ds^2=-(1+2\Phi)dt^2+(1-2\Phi)a^2 \delta_{ij} dx^i dx^j.
\ee
The quantity $\Phi$ is identified as the Newtonian potential corresponding to the perturbation.
Both perturbations $\alpha$ and $\Phi$ are gauge dependent , as pointed out in \cite{Garriga}. However, the quantity
\be\lb{gain}
\zeta=\frac{H}{\dot{\phi}}\alpha+\Phi,
\ee
is not.  Furthermore, this quantity is physically relevant, since it is frozen when the perturbation exit the Hubble horizon and it leads to anisotropies of the CMB. It is then this the quantity of interest to be studied further.

The equations that describe the perturbations described above are obtained by linearization of the Einstein equation, but take a simple form
by introducing the variable \cite{Garriga}
\be\lb{auf}
v=z \zeta,\qquad z=\frac{a\gamma^{3/2}\dot{\phi}}{H}.
\ee
The equation representing the Fourier modes $v_k$ of  (\ref{auf}) then takes the following simple form
\be\lb{pertu}
v_k''+(c_s^2 k^2-\frac{z''}{z})v_k=0.
\ee
Here the derivatives are taken with respect to the conformal time $\tau$. The sound speed for the tachyon model of the previous section is $c_s=1-\gamma_0$, and it is clearly
different than the speed of light.

The task of finding solutions of (\ref{pertu}) for the tachyonic background described above gets complicated by the fact
that the expression for the conformal time $\tau(t)$ as a function of $t$ is not elementary for our tachyon background, and the correspondence between $\tau$ and $t$ it is not to one. For this reason it is convenient
to consider limiting cases. The scale factor $a(t)$ given in (\ref{aL}) goes to the de Sitter regime for large times
$$
a\sim a_0 e^{\sqrt{\frac{\Lambda}{3}}t},\qquad \frac{\gamma_0\sqrt{3\Lambda}}{2} t>>1.
$$ 
For shorter times instead, it has the dependence
$$
a\sim a_0\bigg(\frac{\gamma_0\sqrt{3\Lambda}}{2}\bigg)^{\frac{2}{3\gamma_0}} t^{\frac{2}{3\gamma_0}}\qquad \frac{\gamma_0\sqrt{3\Lambda}}{2} t<<1.
$$
However, the solution of  (\ref{pertu}) requires the knowledge of the dependence of these factors with respect to the conformal time $\tau$ defined through
$$
d\tau=\frac{dt}{a(t)}.
$$
As mentioned above, there is not one to one correspondence between $a(t)$ and $\tau$ for the tachyon solution in consideration. 
However, for short and large times, the correspondence is one to one and is given by
$$
\tau_1=-\frac{\sqrt{3}}{a\sqrt{\Lambda}},\qquad \frac{\gamma_0\sqrt{3\Lambda}}{2} t>>1,
$$
\be\lb{conf}
\tau_2=\frac{2}{a_0\gamma_0\sqrt{3\Lambda}}\bigg(\frac{a}{a_0}\bigg)^{\frac{3\gamma_0-2}{2}},\qquad \frac{\gamma_0\sqrt{3\Lambda}}{2} t<<1.
\ee
The index 1 and 2 has been introduced in order to emphasize that the relation between $\tau$ and $t$ is not bi-univoque.
For the specific case $\gamma_0=3/2$ the power law in $\tau_2$ has to be replaced by a logarithm. 

It is important to remark that, for small values for $a(t)$,  the second (\ref{conf})  implies that the conformal time $\tau$ tends to a large value only when  $\gamma_0\leq 2/3$. For the opposite case, it also takes small values. 
The analysis to be done below is valid for  $\gamma_0\leq 2/3$. 

The slow rolling parameter $\epsilon$ for the tachyon background as a function of the scale factor $a(t)$ is given by
\be\lb{sr}
\epsilon=-\frac{\dot{H}}{H^2}=\frac{3\gamma_0}{2}\frac{1}{1+(\frac{a}{a_0})^{3\gamma_0}}.
\ee
Its behavior in the regimes (\ref{sr}) is the following
$$
\epsilon\sim 0, \qquad \frac{\gamma_0\sqrt{3\Lambda}}{2} t>>1,
$$
\be\lb{comp}
\epsilon\sim\frac{3\gamma_0}{2} \qquad \frac{\gamma_0\sqrt{3\Lambda}}{2} t<<1.
\ee
This suggest that the values of that this parameter takes is always bounded in the evolution of the tachyon background. However, its smallness 
may depend on the value of $\gamma_0$, which is a parameter related to the sound speed $c_s$. Nevertheless, after certain time the slow rolling condition will
be clearly satisfied, and inflation would take place. In order to find the explicit form for the equation (\ref{pertu}) one should take into account that
\be\lb{mt}
\frac{z^{''}}{z}=2 a^2 H^2 (1+\epsilon)=2 a^2 H^2 \bigg\{1+\frac{3\gamma_0}{2}\frac{1}{1+(\frac{a}{a_0})^{3\gamma_0}}\bigg\},
\ee
which is a function of $\tau$ by (\ref{conf}). In addition, the following formula will be useful
$$
2 a^2 H^2 =\frac{2a^2\Lambda}{3}\frac{1+(\frac{a}{a_0})^{3\gamma_0}}{(\frac{a}{a_0})^{3\gamma_0}}.
$$
Starting with the first (\ref{conf}) and taking into account that the scale factor  $a(t)$ takes large values at large times, one leads to 
$$
2 a^2 H^2\sim\frac{2}{\tau_1^2},\qquad \frac{\gamma_0\sqrt{3\Lambda}}{2} t>>1.
$$
Instead, the second (\ref{conf})  together with the fact that $a(t)$ take small values at short times gives that
$$
2 a^2 H^2\sim\frac{2a_0^2\Lambda}{3}(\frac{a_0}{a})^{3\gamma_0-2}\sim \frac{8}{3\gamma_0^2}\frac{1}{\tau_2^2},\qquad \frac{\gamma_0\sqrt{3\Lambda}}{2} t<<1.
$$
Both facts lead to the following behavior
$$
\frac{z''}{z}=\frac{2}{\tau_1^2},\qquad \frac{\gamma_0\sqrt{3\Lambda}}{2} t>>1,
$$
$$
\frac{z^{''}}{z}=\frac{8}{3\gamma_0^2}(1+\frac{3\gamma_0}{2})\frac{1}{\tau_2^2},\qquad \frac{\gamma_0\sqrt{3\Lambda}}{2} t<<1.
$$
In these terms it follows that the equation (\ref{pertu}) to solve takes has the following limiting cases
$$
v_k''+\bigg(c_s^2k^2-\frac{2}{\tau_1^2}\bigg)v_k=0,\qquad \frac{\gamma_0\sqrt{3\Lambda}}{2} t>>1,
$$
\be\lb{eco}
v_k''+\bigg[c_s^2k^2-\frac{8}{3\gamma_0^2}\bigg(1+\frac{3\gamma_0}{2}\bigg)\frac{1}{\tau_2^2}\bigg]v_k=0,\qquad \frac{\gamma_0\sqrt{3\Lambda}}{2} t<<1.
\ee
The two equations (\ref{eco}) are of the same type, but with different numerical parameters. Both can be solved in terms of Henkel functions $H^{(j)}_{\nu_i}(x)$, which represent
the asymptotic behavior in both regimen. This behavior is
$$
v_k\simeq\sqrt{-\tau}\bigg(c_1(k)H^{(1)}_{\nu_1}(-c_s k \tau_1)+c_2(k)H^{(2)}_{\nu_1}(-c_s k \tau_1)\bigg),\qquad \frac{\gamma_0\sqrt{3\Lambda}}{2} t>>1,
$$
\be\lb{aso}
v_k\simeq\sqrt{\tau}\bigg(d_1(k)H^{(1)}_{\nu_2}(c_s k \tau_2)+d_2(k)H^{(2)}_{\nu_2}(c_s k \tau_2)\bigg), \qquad \frac{\gamma_0\sqrt{3\Lambda}}{2} t<<1.
\ee
The values of $\nu_i$ are given by
\be\lb{nues}
\nu_1=\frac{3}{2},\qquad \nu_2=\sqrt{\frac{1}{4}+\frac{8}{3\gamma_0^2}\bigg(1+\frac{3\gamma_0}{2}\bigg)}.
\ee
The constants $c_i(k)$ and $d_i(k)$ for (\ref{aso}) will be defined by the boundary conditions of the problem.
These conditions are characterized  next.

\subsection{Boundary conditions for the modes $v_k$}

Although (\ref{aso}) describes the asymptotic behavior of the solutions for small and large time values, the behavior near the critical time
\be\lb{crit}
 t_c\sim \frac{2}{\gamma_0\sqrt{3\Lambda}}, 
\ee
is not easy to be determined analytically. However (\ref{eco}) suggest that
$$
\frac{z''}{z}=\frac{f(\tau)}{\tau^2},
$$
with $f(\tau)$ a function which takes controlled values, of the order of the values of the parameters $\nu_i$, during the whole the evolution of the tachyonic background. Based on this we may make a crude approximation in order to estimate the actual solution. This approximation consists in extending the region (\ref{eco}) to the critical value $t_c$ defined in  (\ref{crit}).  Although the relation between  $\tau$ and $a$ is not one to one, the relation between $a(t)$ and $t$ is bi-univoque. For this reason, at the time $t_c$ the boundary condition to be imposed is the continuity of $v_k$ and its "time $t$" derivative. This requires to express $\tau$ as a function of $t$ in the regimes considered. The resulting problem resembles a quantum mechanical problem with an step potential.

Consider first a tachyon field with $\gamma_0<2/3$. In this case, for values of $t$ small enough the factor $a(t)$ is also small and $c_s k\tau>>1$ for short wavelengths.  The first (\ref{aso}) then implies that asymptotically
$$
v_k\sim \sqrt{\frac{2}{\pi k}} \bigg(d_1(k)e^{i(k\tau- \frac{\pi \nu_2}{2}-\frac{\pi}{4})}+d_2(k)e^{-i(k\tau- \frac{\pi \nu_2}{2}-\frac{\pi}{4})}\bigg).
$$
By assuming that the solution found should correspond to positive energy plane wave solution
$$
v_k\sim \frac{1}{\sqrt{2k}} e^{ik\tau},
$$
which corresponds to the Bunch-Davies vacuum, it follows that the coefficients $d_i(k)$ are given by 
$$
d_1(k)=\frac{\sqrt{\pi}}{2} e^{i\frac{\pi}{2}(\nu_2+\frac{1}{2})},\qquad d_2(k)=0.
$$
Note that the sign is opposite to many models in the literature, due to the fact that for small times $\tau$ is negative in our case, while is positive for several non tachyon scalar fields.

The other unknown coefficients $c_i(k)$ can be determined by use of the boundary conditions described in the previous paragraphs at $t=t_c$, the result is
$$
c_1(k)=-\frac{d_1(k)}{W}\bigg[ \frac{d(\sqrt{-\tau} H^{(2)}_{\nu_1}(-c_s k \tau))}{d\tau}|_{\tau_1}\sqrt{\tau_2} H^{(1)}_{\nu_2}(c_s k \tau_2)- \sqrt{-\tau_1} H^{(2)}_{\nu_1}(-c_s k \tau_1)\ \frac{d(\sqrt{-\tau} H^{(1)}_{\nu_2}(c_s k \tau))}{d\tau}|_{\tau_2}\bigg],
$$
\be\lb{ger}
c_2(k)=\frac{d_1(k)}{W }\bigg[ \frac{d(\sqrt{-\tau} H^{(1)}_{\nu_1}(-c_s k \tau))}{d\tau}|_{\tau_1}\sqrt{\tau_2} H^{(1)}_{\nu_2}(c_s k \tau_2)- \sqrt{-\tau_1} H^{(1)}_{\nu_1}(-c_s k \tau_1)\ \frac{d(\sqrt{-\tau} H^{(1)}_{\nu_2}(c_s k \tau))}{d\tau}|_{\tau_2}\bigg].
\ee
Here the quantity 
$$
\tau_1= -\frac{\sqrt{3}}{a_0\sqrt{\Lambda}} ,\qquad \tau_2=\frac{2}{a_0\gamma_0\sqrt{3\Lambda}},
$$
has been introduced. The wronskian in the denominator $W$ in (\ref{ger}) is the determinant given by
\[ W= \left| \begin{array}{cc}
\sqrt{-\tau} H^{(1)}_{\nu_1}(-c_s k \tau_1)& \frac{d(\sqrt{-\tau} H^{(1)}_{\nu_1}(-c_s k \tau))}{d\tau}|_{\tau_1}  \\
\sqrt{-\tau} H^{(2)}_{\nu_1}(-c_s k \tau_1) & \frac{d(\sqrt{-\tau} H^{(2)}_{\nu_1}(-c_s k \tau))}{d\tau}|_{\tau_1} \end{array} \right|.\] 
Taking into account the asymptotic behavior
\be\lb{asin2}
H^{(1)}_\nu(x)\sim-H^{(2)}_\nu(x)\sim - \frac{i 2^{\nu}}{\pi}\frac{\Gamma(\nu)}{x^{\nu}},\qquad x<<1,
\ee
it is concluded that for large wavelengths
$$
v_k\sim i(c_2(k)-c_1(k)) \frac{2^{\nu}}{\pi \tau^{\nu-1/2}}\frac{\Gamma(\nu)}{(-c_sk)^{\nu}}.
$$
This solutions differs from the de Sitter one, as expected. The difference is given by the coefficients $c_i(k)$, which are $\gamma_0$  and  $k$ dependent. 

There exist literature \cite{bardo0}-\cite{bardo3} which consider boundary conditions which resemble the ones that have been discussed above. These scenarios are specific inflation models for which some of its particle
contents develop an abrupt mass change during their evolution, or an abrupt change in the state equations. The boundary conditions are imposed on a co-moving surface and are equivalent to the continuity of the Bardeen and the extrinsic curvature on that surface. In our case however, there is no sudden change in the quantities, but it is analogous to model a barrier with large mean slope with an square potential barrier.

By collecting the results described above and the definition $v_k=z \zeta_k$ it follows that
$$
\zeta_k\sim\frac{H}{a \gamma_0^{2}} i(c_2(k)-c_1(k)) \frac{2^{3/2}}{\pi  \tau}\frac{\Gamma(3/2)}{(-c_sk)^{3/2}},
$$
which, in a near de Sitter regime, takes the following form
$$
\zeta_k\sim\frac{\Lambda}{3 \gamma_0^{2}} i(c_2(k)-c_1(k)) \frac{2^{3/2}}{\pi}\frac{\Gamma(3/2)}{(-c_sk)^{3/2}}.
$$
The power spectrum is in this situation 
$$
P_{\zeta}=\frac{1}{2\pi^2}\zeta_k \zeta^\ast_k k^3=\frac{\Lambda^2}{9 \gamma_0^{4}} (c_2(k)-c_1(k))(c_2(k)-c_1(k))^\ast \frac{2}{\pi^3}\frac{\Gamma^2(3/2)}{(c_s)^{3}},
$$
is not $k$ independent in general. This is because the coefficients $c_i(k)$ are functions of $k$ not only in its phase, but in its amplitude. For small $k$ values it follows from (\ref{asin2}) and (\ref{ger}) that
\be\lb{nuest}
P_{\zeta}\sim \frac{1}{k^{\frac{1}{2}+\nu_2}},
\ee
with $\nu_2$ given in (\ref{nues}). Note that this quantity depends on $\gamma_0$, thus its values varies for different tachyons. Thus the solution found shows that for large wavelengths the spectrum differs considerably from the de Sitter one. For large $k$ values the functions $\sqrt{-\tau}H^i_\nu$ behave as a plane wave solution. In this limit, the solution (\ref{ger}) shows that $d_1(k)\sim c_{1}(k)\exp(i\alpha)$ and $c_2(k)\sim 0$, with $\alpha$ a  phase whose value depends on the value of $t_c$. This shows the power spectrum is well represented by the one corresponding to the Bunch-Davies solution, which correspond to the standard inflation model. In fact, the resulting power spectrum is
\be\lb{alm}
P_\zeta\sim \frac{1}{\gamma_0^2}P_{\Lambda},
\ee
thus it is $k$ independent in this approximation, and $P_\Lambda$ is the power spectrum of the standard inflation model. 

In brief, the power spectrum for the present model imitates the standard model for  wavelengths much shorter that the Hubble radius, as shown in (\ref{alm}), it deviates at middle wavelengths and at large wavelengths it has the behavior (\ref{nuest}) which deviate considerably from the standard model.This result coincides with the one get it in the limit of the slow roll approximation, if the power spectrum is a constant the spectral index is $n_s=1$.  

\section{Non gaussianities and three point correlation functions}
The  results found in the previous section seem to support the fact that the power spectrum for this model is $k$ independent. This imitates a gaussian spectrum, although the behavior for very large wave lengths deviates from this spectrum, as stated in (\ref{nuest}). A more specific check for gaussianity is the calculation of the  three point correlation function for the field $\zeta(x,t)$, which is defined in (\ref{gain}). This vanishes in the gaussian case, and its value is a measure of the gaussian deviations. In the presence section, it will be shown that the non gaussianities are completely negligilble for this model.

As discussed in \cite{maldacena}  the definition (\ref{gain})  can be approximated in the slow rolling regime by
\be\lb{gain2}
\zeta=\frac{H}{\dot{\phi}}\alpha+O(\epsilon, \tau)\bigg(\frac{H}{\dot{\phi}}\alpha\bigg).
\ee
The second term of this expression is of higher order, and may be neglected if $\epsilon<1$. Thus, the strategy to be used below is to calculate the three point function for the field $\alpha(x,t)$ namely and to translate it into the three point correlation function $<0(t)|\zeta(x,t)\zeta(y,t)\zeta(z,t)|0(t)>$ by use of (\ref{gain2}). This correlation vanish for gaussian perturbations, thus it is a measure of the deviations of the gaussian spectrum. 
  
One of the best suited formalism for calculating such correlations is the closed path formalism (CPF), which is also useful in non equilibrium physics. Some standard references for their application in cosmology are \cite{calzetta1}-\cite{calzetta6} and also \cite{maldacena}. The general prescription for calculating a mean value of an operator $O(t)$ in this formalism is
 \be\lb{tripon2}
<0|O(x,t)|0>=<0|(\hat{\overline{T}} e^{-i\int_{-\infty(1-i\epsilon)}^t H^I_{int}(t')dt'})O^I(t)(\hat{T} e^{i\int_{-\infty(1+i\epsilon)}^t H^I_{int}(t')dt'})|0>.
\ee
 Here $\hat{T}$ denotes the standard time ordering operation. The hamiltonian for a generic perturbation 
 $$
 H(\alpha, \phi)=\pi \dot{\alpha}-L,\qquad \pi=\frac{\delta L}{\delta \dot{\alpha}},
 $$
should be decomposed as a sum  $H=H_0+H_{int}$ of a free particle hamiltonian $H_0$ plus an interaction term $H_{int}$. The term $H^I_{int}$ defined in (\ref{tripon2}) is the interaction hamiltonian in the interaction picture, namely
 \be\lb{inter}
 H^I_{int}(t)=U_0^{-1}(t,t_0)H_{int}(\alpha(t_0), \pi(t_0), t_0)U_0(t_0, t),
 \ee
 where the free evolution operator $U_0(t,t_0)$ is defined by the following equation and initial conditions
 $$
 \frac{dU(t,t_0)}{dt}=H_0 U(t,t_0),\qquad  U(t_0, t_0)=1.
 $$
 The explicit expression for the  operator $U_0(t, t_0)$ is
 $$
 U_0(t, t_0)=\hat{T}\exp(-i\int_{t_0}^t H_0(t')dt').
 $$
 The interaction operator $O^I(t)$ corresponding to an arbitrary operator $O(t)$ is given through
 $$
 O(t_0)=U^{-1}_0(t, t_0)O^I(t)U_0(t,t_0).
 $$
 In these terms  the three point function to be calculated is given by
\be\lb{tripon}
<0(t)|\alpha(x,t)\alpha(y,t)\alpha(z,t)|0(t)>=<0|(\hat{\overline{T}} e^{-i\int_{-\infty}^t H_I(t')dt'})\alpha_I(x,t)\alpha_I(y,t)\alpha_I(z,t)(\hat{T} e^{i\int_{-\infty}^t H_I(t')dt'})|0>.
\ee
 By assuming that $H_I(t)$ is an small correction, the expression given above can be reduced to
\be\lb{tripon2}
<0(t)|\alpha(x,t)\alpha(y,t)\alpha(z,t)|0(t)>=Re<[-2i\alpha_I(x,t)\alpha_I(y,t)\alpha_I(z,t)\int_{-\infty(1+i\epsilon)}^t H^I_{int}(t')dt']>,
\ee
up to higher order expansion terms. The standard $i\epsilon$ prescription is aimed to cancel out the contributions from far infinity.

In order to apply the formalism described above to the tachyon case, it is convenient to decompose the field $\phi$ as before as a classical solution and a quantum fluctuation $\phi=\phi(t)+\alpha(x,t)$ and to expand the lagrangian
$$
L=\frac{\Lambda(1-\gamma_0)}{\sinh^2\frac{\sqrt{3\gamma_0\Lambda}}{2}\phi}\sqrt{1-\partial_\mu\phi\partial^{\mu}\phi},
$$
up to order three in $\alpha$. The first variation  $\delta^{(1)}L$ is zero due to the equations of motion and the expansion becomes
$$
L=L_0+\delta^{(2)}L+\delta^{(3)}L,
$$
with
$$
\delta^{(2)} L=\frac{1}{a^{\frac{1}{3\gamma_0}}}\bigg[3\phi^2_0\Lambda\sqrt{1-\phi_0^2}\alpha^2+2\sqrt{\frac{3\Lambda}{1-\phi_0^2}}\phi^2_0\dot{\alpha}+\frac{2+\phi_0^2}{(1-\phi_0^2)^{3/2}}\dot{\alpha}^2-\frac{1}{a^2\sqrt{1-\phi^2}} \nabla_i \alpha\nabla^i\alpha\bigg]
$$
and also
$$
\delta^{(3)} L=\frac{1}{a^{\frac{1}{3\gamma_0}}}\bigg[-(3\Lambda)^{3/2}\phi_0^3\sqrt{1-\phi_0^2}\;\alpha^3-\frac{3\Lambda \phi^2_0}{\sqrt{1-\phi_0^2}}\;\dot{\alpha}\alpha^2-\frac{3\sqrt{3\Lambda}\phi_0^3}{(1-\phi_0^2)^{3/2}}\dot{\alpha}^2\alpha
$$
$$
+\frac{1}{\sqrt{1-\phi_0^2}}\alpha \nabla_\mu \alpha \nabla^\mu \alpha+\frac{7\phi_0}{(1-\phi_0^2)^{3/2}}\dot{\alpha} \nabla_\mu \alpha \nabla^\mu \alpha\bigg].
$$
The interaction hamiltonian to be considered is then the following one \footnote{We did not wrote the Taylor expansion of the term $\sinh^2\frac{\sqrt{3\gamma_0\Lambda}}{2}\phi$. However, we had made a calculation considering these terms, which is more cumbersome but do not change the conclusions given below. The omission is for simplicity.}
$$
H_I(t)=\int d^3x \frac{1}{a^{\frac{1}{3\gamma_0}}}\bigg[(3\Lambda)^{3/2}\phi_0^3\sqrt{1-\phi_0^2}\;\alpha^3+\frac{3\Lambda \phi^2_0}{\sqrt{1-\phi_0^2}}\;\dot{\alpha}\alpha^2+\frac{3\sqrt{3\Lambda}\phi_0^3}{(1-\phi_0^2)^{3/2}}\dot{\alpha}^2\alpha
$$
\be\lb{inte}
-\frac{1}{\sqrt{1-\phi_0^2}}\alpha \nabla_\mu \alpha \nabla^\mu \alpha-\frac{7\phi_0}{(1-\phi_0^2)^{3/2}}\dot{\alpha} \nabla_\mu \alpha \nabla^\mu \alpha\bigg].
\ee
The field $\alpha^I(x,t)$ is assumed to be a quantum one, and may be expanded as
$$
\alpha^I(x,t)=\int \frac{d^3k}{(2\pi)^3}\bigg[\alpha_k(t)e^{ikx} a_k+\alpha^{\ast}_k(t)e^{-ikx}a^{\dag}_k\bigg],
$$
where $a_k$ are operators that annihilates the past asymptotic vacuum 
$$
a_k|0>=0,
$$ 
and both $a_k$ and $a_k^\dag$ realize the standard harmonic oscillator algebra
$$
[a_k,a^\dag_{k'}]=\delta_{kk'}.
$$
The behavior of the functions $\alpha_k(t)$ can be inferred by results of the previous section in the approximation (\ref{gain}). This behavior is described by
\be\lb{bijo}
\alpha(k,\tau)\to -\frac{H\tau}{\gamma_0^{3/2}}\sqrt{\frac{2}{c_s k}} d_1(k)e^{ik c_s \tau},\qquad -k c_s \tau>>1
\ee
\be\lb{bijo2}
\alpha(k,\tau)\to \frac{H}{\gamma_0^{3/2}}\frac{2i}{(2c_s k)^{\frac{3}{2}}} (c_1(k)-c_2(k)),\qquad -k c_s \tau<<1
\ee
The next task is to show that the three point correlation function vanish for our model. Consider first the contribution from the terms of the hamiltonian proportional to $\alpha^3(t)$.
The contribution to the correlation function in momentum space is
$$
<\alpha(k_1,0)\alpha(k_2,0)\alpha(k_3,0)>_1= Re<2i \alpha(k_1,0)\alpha(k_2,0)\alpha(k_3,0)
$$
$$
\int_{-\infty(1+i\epsilon)}^0 d\tau' \int d^3x(-H\tau')^{1/3\gamma_0}\alpha(x,\tau')\alpha(x,\tau')\alpha(x,\tau')>
$$
$$
= -Re<2i \alpha(k_1,0)\alpha(k_2,0)\alpha(k_3,0)\int \;d^3x \;e^{i(q_1+q_2+q_3)\cdot x}\int \frac{d^3q_1}{(2\pi)^3} \frac{d^3q_2}{(2\pi)^3}\frac{d^3q_3}{(2\pi)^3}
$$
$$
\int_{-\infty(1+i\epsilon)}^0 d\tau' \;(-H\tau')^{1/3\gamma_0}\alpha(q_1,\tau')\alpha(q_2,\tau')\alpha(q_3,\tau') >
$$
$$
= -Re\bigg[ 2i \alpha(k_1,0)\alpha(k_2,0)\alpha(k_3,0))\int \;d^3x \;e^{i(q_1+q_2+q_3)\cdot x}\int \frac{d^3q_1}{(2\pi)^3} \frac{d^3q_2}{(2\pi)^3}\frac{d^3q_3}{(2\pi)^3}
$$
$$
\int_{-\infty(1+i\epsilon)}^0 d\tau' \;(-H\tau')^{1/3\gamma_0}\alpha^\ast(k_1, \tau' )\alpha^\ast(k_2, \tau' )\alpha^\ast(k_3, \tau' )(2\pi)^9 \delta(q_1-k_1)\delta(q_2-k_2)\delta(q_3-k_3) \bigg]
$$
$$
= -Re\bigg[ 2i\alpha(k_1,0)\alpha(k_2,0)\alpha(k_3,0)\delta(q_1+q_2+q_3)
\int_{-\infty(1+i\epsilon)}^0 d\tau' \;b (-H\tau')^{1/3\gamma_0}\alpha^\ast(k_1, \tau' )\alpha^\ast(k_2, \tau' )\alpha^\ast(k_3, \tau' )\bigg].
$$
In the third step, the standard Wick theorem was used.
The integral inside of the last expression can be approximated by taking into account (\ref{bijo})-(\ref{bijo2}) as follows
$$
\int_{-\infty(1+i\epsilon)}^0 d\tau' \;b (-H\tau')^{1/3\gamma_0}\alpha^\ast(k_1, \tau' )\alpha^\ast(k_2, \tau' )\alpha^\ast(k_3, \tau' )\sim \tau^{1+\frac{1}{3\gamma_0}}E_{\frac{1}{3\gamma_0}}(k c_s \tau)|_{-\infty(1+i\epsilon)}^{0}.
$$
with $E_{\frac{1}{3\gamma_0}}(x)$ the exponential integral function. The prescription $+i\epsilon$ cancels the contribution from the asymptotic past $\tau\to-\infty$. The limit of  $\tau^{1+\frac{1}{3\gamma_0}}E_{\frac{1}{3\gamma_0}}(k c_s \tau)$ when $\tau\to 0$  also vanish. Thus the correlation terms
$$
\int_{-\infty(1+i\epsilon)}^0 d\tau' \;b (-H\tau')^{1/3\gamma_0}\alpha^\ast(k_1, \tau' )\alpha^\ast(k_2, \tau' )\alpha^\ast(k_3, \tau' )\sim 0.
$$
The same type of calculation also shows that  contributions from the other terms are approximately zero, the integrals are different but the same behavior is obtained in the limit of integrations. Thus the tachyon correlation function seems to be well approximated by a gaussian one. As we expect, the small parameters which are required for slow roll inflation with the sight magnitude of density perturbations, lead to primordial density perturbations which are Gaussian to a very high degree of accuracy.

\section{Summaries and discussions}
In the present work, we have studied a particular tachyon inflationary model. In the slow-roll approximation we have found the parameters $\epsilon$ and $\eta$. With this parameters we have also obtained the spectrum index $n_s$ as a function of the cosmic time $t$. We have plotted $n_s(t)$ for different values of the barotropic index, for example $\gamma_0=0.004$, in this case the spectral index at a initial value of time is $n_s=0.964$, in agreement with the data released by Planck Mission \cite{Planck2015}. We could see that the spectral index function increase slowly and goes to the unity, that correspond to a the Sitter stage.
We have found that the spectrum for short wavelengths is qualitatively the same as the standard inflation model. However, we checked this by calculating the three point gaussian correlation function for the primordial scalar fluctuation in the model, and we have found that is close to zero, thus supporting the gaussianity of the spectrum.

The new parameter of the model we have considered is $\gamma_0$, which is related to the speed of sound $c_s$. The analysis made in the present work is for $\gamma_0<2/3$. There is an apparent problem for extending this analysis to the other regime of the parameter, which is the fact that the conformal time in this case tends to zero for asymptotic coordinate times, and therefore one can not impose the standard Bunch-Davies conditions adapted to this case. One may extend our results to the complementary parameter space, but this is equivalent to assume that the behavior of the model is continuous with respect to the choice of this parameter.  We ignore if there is a phase transition when crossing the value $\gamma_0=2/3$. Perhaps for higher values the non linear effects of the tachyon model may change qualitatively the behavior of the background. We leave this as a part of a future investigation.

\section*{Acknowledgments}
The authors are supported by the CONICET, Argentina.



\begin{thebibliography}{99}
\bibitem{Staro} A. A. Starobinsky, Phys. Lett. \textbf{99}, 24 (1980); A. H. Guth, Phys. Rev. D \textbf{23}, 347 (1981); A. D. Linde,Phys. Lett. B \textbf{108}, 389 (1982); A. Albreht and P. J. Steinhardt, Phys. Rev. Lett. \textbf{48}, 1220 (1982).

\bibitem{Lyth} D. H. Lyth and A. R. Liddle, \textit{The Primordial Density Perturbation} (Cambridge University Press, 2009).

\bibitem{TaS}
A. Sen, JHEP 0204, 048 (2002); JHEP 0207, 065 (2002); Mod. Phys. Lett. A 17, 1797 (2002);
A. Sen, JHEP 9910, 008 (1999); M. R. Garousi, Nucl. Phys. B584, 284 (2000); JHEP 0305, 058 (2003); E. A. Bergshoeff, M. de Roo, T. C. de Wit, E. Eyras, S. Panda, JHEP 0005, 009 (2000); D. Kutasov and V. Niarchos, Nucl. Phys. B 666, 56 (2003).

\bibitem{Pad}
T. Padmanabhan, Phys. Rev. D \textbf{66}, 021301 (2002).

\bibitem{FE}
A. Feinstein, Phys.Rev.D \textbf{66}, 063511 (2002).

\bibitem{Nozari} K. Nozari and N. Rashidi, Phys. Rev. D, \textbf{88}, 023519 (2013).

\bibitem{Dvali} G. R. Dvali and S.-H. H. Tye, Brane inflation, Phys. Lett. B \textbf{450}, 72 (1999).

\bibitem{Sami}  M. Sami, P. Chingangbam and T. Qureshi, Phys. Rev. D \textbf{66}, 043530, (2002).

\bibitem{Abramo}
L. R. W. Abramo and F. Finelli, Phys. Lett. B \textbf{575}, 165 (2003).

\bibitem{Campo}
S. del Campo, R. Herrera, and A. Toloza, Phys. Rev. D \textbf{79}, 083507 (2009).

\bibitem{Jain}
R. K. Jain, P. Chingangbam, and L. Sriramkumar, Nucl. Phys. B \textbf{852}, 366 (2011).

\bibitem{Calcagni}
G. Calcagni, and A. R. Liddle, Phys. Rev. D \textbf{74}, 043528 (2006).

\bibitem{LuisTaq2}
L. P. Chimento, M. Forte, G.M. Kremer, M. O. Ribas, Gen. Rel. Grav. \textbf{42}, 1523-1535, (2010) [arXiv:0809.1919v2].

\bibitem{SanSan} Iv\'{a}n E. S\'{a}nchez G., Osvaldo P. Santill\'{a}n, Gen. Rel. and Grav. \textbf{47}:118 (2015) [arXiv:1502.01060].










\bibitem{Gibbons} G. W. Gibbons, Phys. Lett. B, \textbf{537}, 1, (2002).

\bibitem{Lyth1} D. H. Lyth and A. Riotto, Phys. Rev. Lett. \textbf{97},  121301 (2006), [arXiv:astro-ph/0607326].

\bibitem{LuisTE}
L. P. Chimento, Phys. Rev. D \textbf{69}, 123517 (2004).

\bibitem{ISG}
Iv\'{a}n E. S\'{a}nchez G., Physical Review D \textbf{90}, 027308 (2014).

\bibitem{Malda} N. Lambert, H. Liu and J. Maldacena, JHEP 0703, 014 (2007), [arXiv:hep-th/0303139].

\bibitem{Crema} Daniel Cremades, Fortsch. Phys. 54, 357-365 (2006), [arXiv:hep-th/0512294].

\bibitem{Leblond} Louis Leblond, Sarah Shandera, JCAP 0701, 009 (2007) [arXiv:hep-th/0610321].

\bibitem{silver}  M. Alishahiha, E. Silverstein and D. Tong Phys. Rev. D 70 (2004) 123505.

\bibitem{silver2} E. Silverstein and D. Tong Phys. Rev. D 70 (2004) 103505.

\bibitem{Cook} J. L.Cook, L. Sorbo, Phys. Rev. D \textbf{85}, 023534 (2012).

\bibitem{Planck2015} P. A. R. Ade \emph{et al}, [arXiv:1502.02114v1].

\bibitem{Garriga} Jaume Garriga, V.F. Mukhanov, Phys. Lett. B \textbf{458}: 219-225, (1999).

\bibitem{Muk}
V. F. Mukhanov, H. A. Feldman, and R. H. Brandenberger, Phys. Rep. \textbf{215}, 203 (1992).

\bibitem{MukLibro}
V. Mukhanov, \textit{Physical Foundations of Cosmology}, Cambridge University Press, Cambridge, 2005.

\bibitem{bardo0}  H. Firouzjahi and S. Khoeini-Moghaddam JCAP 1102 (2011) 012.
\bibitem{bardo1} I. Zaballa and M. Sasaki, ÒBoosted perturbations at the end of inflation,Ó arXiv:0911.2069.
\bibitem{bardo2} D. H. Lyth, K. A. Malik, M. Sasaki and I. Zaballa,  JCAP 0601, 011 (2006).
\bibitem{bardo3} I. Zaballa, A. M. Green, K. A. Malik and M. Sasaki JCAP 0703, 010 (2007).

\bibitem{maldacena} J. Maldacena JHEP 0305 (2003) 013.

\bibitem{calzetta1} E. Calzetta and B. L. Hu, Phys. Rev. D35, 495 (1987).
\bibitem{calzetta2}  E. Calzetta and B. L. Hu, Phys. Rev. D40, 656 (1989).
\bibitem{calzetta3}  E. Calzetta and B. L. Hu, Phys. Rev. D37, 2878 (1988).

\bibitem{calzetta4} E. Calzetta, Ann. Phys. (N.Y.) 190, 32 (1989)

\bibitem{calzetta5}  B. L. Hu, ÒQuantum Statistical Processes in the Early UniverseÓ in Quantum Physics
and the Universe, Proc. Waseda Conference, Aug. 1992 ed. M. Namiki et al (Pergamon
Press, Tokyo, 1993). Vistas in Astronomy 37, 391 (1993)

\bibitem{calzetta6}  B. L. Hu, J. P. Paz and Y. Zhang ÒQuantum Origin of Noise and Fluctuations in
CosmologyÓ, in The Origin of Structure in the Universe, edited by E. Gunzig and P.
Nardone (Kluwer, Dordrecht, 1993), p. 227.





\end{thebibliography}
\end{document}